\documentclass[%
 reprint,
 superscriptaddress,
 amsmath,amssymb,
 aps,
 prl,
]{revtex4-2}

\usepackage{graphicx}
\usepackage{dcolumn}
\usepackage{bm}
\usepackage{soul}
\usepackage{comment}

\usepackage{gensymb}
\usepackage[usenames,dvipsnames]{xcolor}
\newcommand{\pipj}[2]{\frac{\partial #1}{\partial #2}}

\begin{document}

\preprint{APS/123-QED}

\title{Motility Modulates the Partitioning of Bacteria in Aqueous Two-Phase Systems}

\author{Jiyong Cheon}
\thanks{Equal contribution}
\affiliation{Department of Physics, Ulsan National Institute of Science and Technology, Ulsan, Republic of Korea}

\author{Kyu Hwan Choi}
\thanks{Equal contribution}
\affiliation{Department of Chemical Engineering, University of California, Santa Barbara, Santa Barbara, CA, USA}
\affiliation{Department of Chemical Engineering, Stanford University, Stanford, CA, USA}

\author{Kevin J. Modica}
\affiliation{Department of Chemical Engineering, University of California, Santa Barbara, Santa Barbara, CA, USA}

\author{Robert J. Mitchell}
\affiliation{Department of Biological Sciences, Ulsan National Institute of Science and Technology, Ulsan, Republic of Korea}

\author{Sho C. Takatori}
\email{stakatori@stanford.edu}
\affiliation{Department of Chemical Engineering, Stanford University, Stanford, CA, USA}

\author{Joonwoo Jeong}
\email{jjeong@unist.ac.kr}
\affiliation{Department of Physics, Ulsan National Institute of Science and Technology, Ulsan, Republic of Korea}

\date{\today}

\begin{abstract}
We study the partitioning of motile bacteria in an aqueous two-phase mixture of dextran (DEX) and polyethylene glycol (PEG), which can phase separate into DEX-rich and PEG-rich phases.
While non-motile bacteria partition exclusively into the DEX-rich phase in all conditions tested, we observed that motile bacteria penetrate the soft DEX/PEG interface and partition variably among the two phases.
For our model organism \textit{Bacillus subtilis}, the fraction of motile bacteria in the DEX-rich phase increased from 0.58 to 1 as we increased DEX composition within the two-phase region.
We hypothesized that the chemical affinity between DEX and the bacteria cell wall acts to weakly confine the bacteria within the DEX-rich phase; however, motility can generate sufficient mechanical forces to overcome the soft confinement and propel the bacteria into the PEG-rich phase. 
Using optical tweezers to drag a bacterium across the DEX/PEG interface, we demonstrate that the overall bacteria partitioning is determined by a competition between the interfacial forces and bacterial propulsive forces.
Our measurements are supported by a theoretical model of dilute active rods embedded within a periodic soft confinement potential.
\end{abstract}

\maketitle


At thermodynamic equilibrium, Boltzmann statistics governs how particles are distributed in an energy landscape, but it does not apply to motile particles at non-equilibrium~\cite{albertsson1970a, doi2013, Knippenberg2024}. For passive colloids dispersed in a liquid-liquid phase-separated system, such as water and oil, their final partitioned location is determined by surface properties and interfacial tension ~\cite{Byun2018}.
When interfacial tension between liquid phases dominates over the difference in particle-phase interactions, the interface tends to form the lowest energy state that colloids are adsorbed at the interface~\cite{wu2016}. Otherwise, the colloids condense into their preferred bulk phase, following a Boltzmann distribution~\cite{Byun2018, singh2018, mastiani2019}. For micron-sized particles, the energy difference between bulk and interfacial states far exceeds $k_{\text{B}} T$, even with low interfacial tensions, $10^{-5}$~N/m~\cite{Atefi2014}. As such, passive particles reliably condense into equilibrium ground states.

In contrast, motile particles, such as microorganisms, generate nonequilibrium forces through their motility, which allows them to overcome energy barriers and deviate from equilibrium distributions~\cite{Knippenberg2024, caprini2019}. This raises fundamental questions about their partitioning behavior in phase-separated systems, where interfacial surface tension poses a significant barrier.

An aqueous two-phase system (ATPS) of dextran (DEX) and polyethylene glycol (PEG) can exhibit a thermodynamically stable liquid-liquid phase separation consisting of PEG-rich and DEX-rich phases~\cite{Atefi2014}.  
Prior works have reported the partitioning behavior of non-motile biological cells and passive colloidal particles into the DEX-rich and PEG-rich phases~\cite{Albertsson1970, BAIRD1961, Hofsten1966, ma2020cell, huang2021characterization, sakuta2019aqueous, Byun2018}. 
However, the fundamental understanding of the nonequilibrium partitioning and barrier-crossing behavior of motile colloids in ATPS is still lacking, despite the ubiquity of living bacteria in many biological and ecological multiphase systems \cite{yoshizawa2020biological, iqbal2016aqueous, dwidar2013patterning}.
Given the strong nonequilibrium forces generated through their motility, motile bacteria may overcome the interfacial surface tension at the two-phase boundaries and assume a highly non-Boltzmann distribution.

\begin{figure}
\centering 
\includegraphics{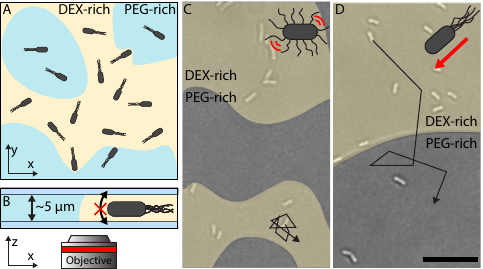}
    \caption{
    Motile bacteria \textit{B. subtilis} immersed in an aqueous two-phase system (ATPS) of dextran (DEX) and polyethylene glycol (PEG) partition variably across the two phases compared to non-motile ones, which partition exclusively into the DEX-rich phase. 
    A) Top and B) side views of our quasi-2D setup for measuring bacterial partitioning. The chamber thickness is $5 \mu$m to prevent the bacteria (average body length $\approx 5 \mu$m) from orienting vertically.
    C) Non-motile \textit{B. subtilis} partition exclusively into the DEX-rich phase (yellow region), whereas D) motile ones partition into both phases. 
    Total concentrations of DEX and PEG are 3.2 and 2.5 wt/wt\%, respectively. The scale bar is 20$\mu$m.}
    \label{fig:microscopy}
\end{figure}

In this Letter, we combine experiment and theory to study the partitioning of motile bacteria in a DEX/PEG system using \textit{Bacillus subtilis} as a model organism.
As shown in Fig.~\ref{fig:microscopy}A-B, we confined the bacteria into a quasi-2D system immersed in a phase-separated mixture of PEG-rich domains surrounded by DEX-rich phase. 
We observed that non-motile \textit{B. subtilis} exhibited Brownian motion and partitioned entirely into the DEX-rich phase (Fig.~\ref{fig:microscopy}C).
In contrast, motile \textit{B. subtilis} distributed more evenly across the two phases and reached a dynamic steady-state, with cells frequently crossing the DEX/PEG interface~(Fig.~\ref{fig:microscopy}D; see also Movie S1).
We also observed an asymmetric interfacial deformation of bacteria penetrating the two-phase boundary, biased toward the DEX-rich phase:
when crossing from DEX $\rightarrow$ PEG, the interface bends \textit{away} from the bacterium, generating a force against bacterial swimming. 
In contrast, when crossing from PEG $\rightarrow$ DEX, the interface bends \textit{towards} the bacteria, generating a force aligned with swimming.
Based on these observations, we hypothesized that chemical affinities between dextran sugars in the buffer and the chemical components on the \textit{B. subtilis} cell wall (such as other sugars) act to weakly confine the bacterium within the DEX-rich phase; however, motility can generate sufficient mechanical forces to overcome this `soft' confinement and propel the bacterium into the PEG-rich phase.
We further hypothesize that the partitioning of motile bacteria in the two phases is determined by a competition between chemical affinity and motility forces. 

To quantify bacterial partitioning, we defined the partitioning ratio as the number of bacteria in the DEX-rich phase divided by the total number of bacteria in the sampled area~(see Supplemental Figs.~S2 and S3), $n_{\mathrm{DEX}}/(n_{\mathrm{DEX}}+n_{\mathrm{PEG}})$.
As shown in Fig.~2, non-motile \textit{B. subtilis} has a partitioning ratio of 1 across all DEX concentrations tested (1.5 - 8.0 wt/wt\%) with a fixed PEG concentration (2.5 wt/wt\%). 
We confirmed that the non-motile bacteria are not kinetically trapped in the DEX-rich phase by agitating the system with turbulent mixing, and also initializing the bacteria in the PEG-rich phase. Despite all attempts to force non-motile bacteria into the PEG-rich phase, every protocol yielded 100\% partitioning into the DEX-rich phase at steady state.
In contrast, the partitioning ratio for the motile bacteria increased monotonically from $0.58 \pm 0.10$ at 1.5 wt/wt\% DEX concentration, to 1.00 at 8.0 wt/wt\%.  

Several possible mechanisms may explain our biased partitioning of motile bacteria, including motility, chemotaxis, and chemical affinity.
One possible mechanism is based on the difference in viscosities of the PEG-rich and DEX-rich phases. 
A higher viscosity in the DEX-rich phase may slow down the bacterium preferentially in the DEX-rich phase, leading to its accumulation in the DEX-rich phase. 
This effect is analogous to active Brownian particles' motility-induced phase separation (MIPS), where slower particles trigger clustering~\cite{Cates2015, van2019interrupted, caporusso2020motility}.
Variable swimming speeds would produce a density distribution as $n(x) \sim 1/U(x)$, and we would need to have speeds in the DEX-rich phase to be 72\% slower ($U_{\mathrm{DEX}}/U_{\mathrm{PEG}}= 0.28$) to match the partitioning observed in our experiments: 
$n_{\mathrm{DEX}}/(n_{\mathrm{DEX}}+n_{\mathrm{PEG}}) = (1+U_{\mathrm{DEX}}/U_{\mathrm{PEG}})^{-1} = 0.78 \pm 0.08$, for 3.2 wt/wt\% DEX concentration.
However, in a separate control experiment, we independently measured $U_{\mathrm{DEX}} = 30.7 \pm 6.1~\mathrm{\mu m/s}$ and $U_{\mathrm{PEG}} = 34.5 \pm 6.4~\mathrm{\mu m/s}$ under the same DEX concentrations. These measurements indicate that the speed in the DEX-rich phase is only 11\% slower ($U_{\mathrm{DEX}}/U_{\mathrm{PEG}}= 0.89$).
We therefore ruled out variable swimming speeds and viscous drag as mechanisms for our observed partitioning.

Chemotaxis is another possible mechanism that may explain the preferential distribution of bacteria in the DEX-rich phase, as \textit{B. subtilis} is known to respond to sugar (glucose) gradients~\cite{Garrity1995}. 
Upon imaging the DEX-PEG interface using FITC-tagged DEX (see Supplemental Fig.~S4), we observed a rapid decay in DEX-FITC concentration over $\approx 2~\mathrm{\mu m}$. Since chemotaxis requires chemical gradients to persist over distances larger than the run length ($l_{\mathrm{run}}$)~\cite{tindall2008overview} ($l_{\mathrm{run}} \gg 100~\mathrm{\mu m}$ in our system; See Supplementary Information and Fig.~S7), chemotaxis is unlikely to explain the biased partitioning of motile bacteria in our experiments.

\begin{figure}[t]
\centering 
\includegraphics{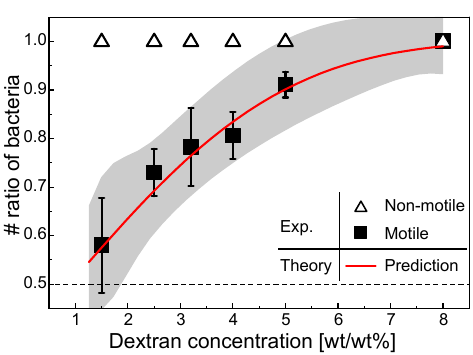}
    \caption{
     Experimental and predicted partitioning ratios of bacteria in DEX-rich phase versus dextran concentration of the overall mixture.
     Open and filled symbols are experimental results of non-motile and motile bacteria, respectively, and the red solid curve is the predicted partitioning ratio based on our theory, Eqs.~\ref{eq:1DAdvectionDiffusion} and \ref{eq:partition}, fitted to two adjustable parameters. The shaded region represents the 95\% confidence band of our model. The horizontal dashed line at 0.5 represents a theoretical partitioning ratio when the two-phase boundary vanishes at the critical composition.
     }
\label{fig:partitioning result}
\end{figure}

\begin{figure}[t]
\centering
\includegraphics{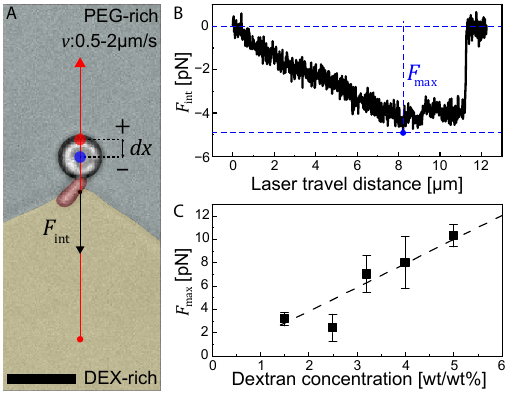}\caption{Optical tweezers were used to drag a single bacterium across the two-phase boundary and measure the forces imparted on the bacterium. A) Representative snapshot of a bacterium attached to a tracer bead crossing the interface. The red solid arrow is the path of the optical trap. The force generated by the interface, $F_\text{int}$, is calculated by measuring the displacement, $dx$, between the laser focal point (red dot) and the tracer bead center (blue dot). Negative values indicate the force is directed against the trap movement. The scale bar is 5~$\mu$m. B) Representative force profile of $F_\text{int}$ as the bacterium crosses the two-phase boundary. The laser travel distance is set to zero when the cell body begins to cross the DEX/PEG boundary. C) Maximum forces ($F_\text{max}$) were measured for various DEX concentrations, and the dashed line represents a linear regression for parameter estimation in our theory, Eq.~\ref{eq:1DAdvectionDiffusion}.}
    \label{fig:force&energy curve}
\end{figure}

To these ends, we hypothesized that the chemical affinity between the DEX-rich phase and the sugars and amino acids of peptidoglycan surrounding the bacteria~\cite{DZIARSKI19942100} drives the partitioning.
We performed optical tweezer experiments to measure the forces required to drag single bacterial cells across the DEX-PEG two-phase boundary (Fig.~\ref{fig:force&energy curve}A).
We attached a tracer silica bead as a calibration handle to a non-motile bacteria surface using click chemistry~(See Fig.~\ref{fig:force&energy curve}A, Supplemental Figs.~S5 and Movie S2).
We translated the trap location from the DEX-rich phase into the PEG-rich phase at small speeds (0.5 or 2$\mu m/s$) to reduce the viscous drag force on the bead.
Using the displacement of the bead center from the trap center, we measured the force generated by the two-phase boundary on the bacterium, $F_\mathrm{int}$, as shown in Fig.~\ref{fig:force&energy curve}B and Supplemental Fig.~S5.

As the cell body crossed the interface from DEX-rich to PEG-rich phases, the interface deformed and pulled the bacterium back towards the DEX-rich phase, causing the force to increase as a function of penetration distance.
When the force on the bacterium by the interface, $F_\text{int}$, reached its maximum, $F_\text{max}$, the interface maintained a constant contact angle and started slipping along the surface of the dragged bacterium (See Fig. S5 and S.I.). When the bacterium completely crossed the interface, the interface recovered its initial unperturbed shape.

We conducted the measurement across several different dextran compositions in the ATPS mixture. 
As shown in Fig.~\ref{fig:force&energy curve}C, $F_\text{max}$ ranges from 1 to 10 pN and increases with dextran concentration. 
These force measurements are direct reporters of the force experienced by swimming bacteria as they cross the interface from the DEX-rich to PEG-rich phases. 
The contact angle formed by the triple point of the DEX/PEG/bacterium interface was the same between freely swimming bacteria and dragged bacteria in the optical tweezers (See Movie~S4 and Fig.~S6). Our measurements corroborate our hypothesis that thermal forces alone are not sufficient to allow non-motile bacteria to cross even the weakest two-phase boundary.For a $1 \mu m$ colloid, thermal forces are three orders of magnitude lower than the softest DEX/PEG interface in our system, $k_\text{B}T/1~\mathrm{\mu m} \approx 10^{-3}~\mathrm{pN} \ll 1~\mathrm{pN}$.
Therefore, in the absence of a propulsion force, all our non-motile \textit{B. subtilis} partitioned into the DEX-rich phase across all conditions tested in Fig.~\ref{fig:partitioning result}. 
In contrast, the propulsion forces of motile bacteria, $F_\text{prop} \approx$10~pN~(See Fig.~S7C and \cite{cheon2024}), are sufficiently large to overcome the maximum forces generated by the interface, $F_\text{max}$.

At lower DEX concentrations, the maximum forces $F_\text{max}$ are smaller (see Fig.~\ref{fig:force&energy curve}C) and the propulsion forces, $F_\text{prop}$ are large enough to enable the bacteria to swim unimpeded between phases. This leads to a partitioning ratio near 0.5, as shown in Fig.~\ref{fig:partitioning result}.
For the largest DEX concentration tested, 8.0 wt/wt\%, the interfacial forces ($F_\text{int}$) are too strong for the laser trap to drag the bacteria completely inside the PEG-rich phase.
Namely, the laser trap loses control of the colloid prior to crossing the interface and we are unable to measure $F_\text{max}$. 
Consistent with this observation, the motile \textit{B. subtilis} partitioned completely inside the DEX-rich phase at a 8.0 wt/wt\% DEX concentration because the propulsion forces were always smaller than the interfacial barrier, $F_\text{prop} < F_\text{max}$ (See Movie S3).

To corroborate our experimental measurements, we modeled the affinity of the bacteria to the DEX-rich phase as an external potential field that drives the bacteria to regions of high DEX concentration.
We modeled the bacteria as a collection of dilute, thin, rod-shaped, active Brownian particles in two-dimensions whose motion is described via the Smoluchowski equation, $\frac{\partial P}{\partial t} + \nabla \cdot  \left( \frac{P}{\zeta} \mathbf{F}  \right) + \frac{\partial}{\partial \theta} \left( \frac{P}{\zeta_R} (L_\text{B} + L_\text{ext}) \right) = 0$. 
The dilute (i.e. no interparticle interactions) rod probability distribution, $P(\mathbf{x},t)$, is governed by the forces $\mathbf{F} = \mathbf{F}_\text{prop} + \mathbf{F}_\text{B} + \mathbf{F}_\text{ext}$.
The Brownian contributions are $\mathbf{F}_\text{B} = -k_{\mathrm{B}}T\nabla \ln P$ and $L_\text{B} = -k_{\mathrm{B}}T\frac{\partial \ln P}{\partial \theta}$. The active force is a polar force directed along the rod's semi-major axis at angle $\theta$ and written as $\mathbf{F}_\text{prop} = F_\text{prop} (\cos{\theta} \mathbf{e}_x + \sin{\theta} \mathbf{e}_y)$. The external forces and torques are integrated along the rod body and act on the center of mass to give the expressions $\mathbf{F}_\text{ext} = - \nabla V$ and $L_\text{ext}=- \frac{\partial V}{\partial \theta}$. To gain analytical tractability, we ignore the influence of surface tension between the DEX-rich and PEG-rich phases on the bacterial motion; additionally, we treat the bacterium as a thin rigid rod of negligible diameter.

While the PEG-rich liquid domains are often curved and heterogeneous in size, we modeled this system as a planar repeating lamellar phase of alternating DEX-rich and PEG-rich phases that are periodic in the $x$-direction and uniform in the $y$-direction. The affinity of the bacterium to reside in each phase is described as a 1D periodic potential field along the $x$-axis, with no gradients in the $y$-axis. This construction simplifies the problem by constraining the partial differential equation to depend on two independent variables ($x$ and $\theta$). The Smoluchowski equation is then simplified to~\cite{D3SM01340E,BaskaranPRE2008} 
\begin{multline}
\label{eq:1DAdvectionDiffusion}
    \pipj{P}{t}+\pipj{}{x} \left[\frac{P}{\zeta}\left( F_\text{prop} \cos{\theta} - k_\mathrm{B}T \pipj{\ln P}{x}  - \pipj{V}{x}\right) \right]\\ +  \pipj{}{\theta} \left[ P\left( - \frac{1}{\tau_\text{R} }\pipj{\ln P}{\theta}  - \frac{1}{\zeta_R}\pipj{V}{\theta} \right)\right] = 0,
\end{multline}
where $x$ and $\theta$ are the position and orientation of the active particle, $\zeta$ and $\zeta_R$ are translational and rotational drag coefficients, $F_\text{prop}$ is the propulsion force, $\tau_\text{R}$ is the reorientation time. The potential energy $V(x,\theta)$ encodes the energetic cost for the bacterium to move from the DEX-rich phase to the PEG-rich phase (see Supplementary Information for the functional form). Assuming the bacterium is a thin rod, we can relate their rotational and translational drag coefficients using the length of the bacterium $\zeta/\zeta_R = 6/\ell_\text{bact}$. 
We keep the reorientation time ($\tau_\text{R}$) general as it may arise from non-thermal reorientation mechanisms like flagella unbundling.
For simplicity, we ignore the orientation dependence on the translational drag coefficient and treat the PEG-rich and DEX-rich phases as a 1D lamellar system with equal domain width $L$ in a periodic system of size $2L$. 
This model can be described with five parameters: the ratio of the maximum interface restoring force to the propulsion force $F_\text{max}/ F_\text{prop}$, the ratio of propulsion force relative to thermal energy $F_\text{prop} L/( k_\text{B}T)$, the ratio of two time-scales, thermal diffusivity relative to the reorientation time $L^2 \zeta/(k_\text{B}T \tau_\text{R})$, the scaled bacteria length $\ell_\text{bact}/L$, and the scaled effective interface width $\delta / L$. $\delta$ is the length-scale over which the bacterium interacts with the interface from the PEG-rich to DEX-rich phases.

This model assumes that the bacterial partitioning is determined solely by the balance between the propulsion force, Brownian motion, and the potential energy field $V(x,\theta)$. The current model does not consider the impact of interfacial tension and interface deformation. We expect the interaction energy field $V(x,\theta)$ will exert a force acting against bacteria motion that is roughly the same magnitude going from a PEG-rich phase to a DEX-rich phase or back. Therefore, the work needed to deform the interface should not bias the partitioning on one side or the other and should only influence the dynamics of the model as it approaches a steady state.

Based on this theoretical model, we solved the steady-state Smoluchowski equation using the methods described in a recent work \cite{D3SM01340E}. The distribution of bacteria in space and orientation is converted into the familiar partitioning ratio by integrating the probability over all orientations (bacteria number density $n(x,t) = \int P(x,\theta) d\theta$) and then integrating that density over the lower energy region corresponding to the DEX-rich phase ($\Omega_{\mathrm{DEX-rich}}$):
\begin{align} \label{eq:partition}
 \frac{n_{\mathrm{DEX}}}{n_{\mathrm{DEX}}+n_{\mathrm{PEG}}} = \int_0^{2\pi} \int_{\Omega_{\mathrm{DEX-rich}} }P(x,\theta) \ dx \ d\theta.
\end{align}

Due to the large parameter space, we constrain the experimentally determined parameters of the propulsion force
$F_\text{prop} = 10~$pN, bacteria length, $\ell_\text{bact} =  5~\mu$m, average domain diameter $L = 200~\mu$m, and thermal energy $k_\text{B}T = 4 \times 10^{-3}~$pN$\cdot\mu$m, which specify $F_\text{prop} L/ (k_\text{B}T) = 5 \times 10^5$ and $\ell_\text{bact}/L = 0.025$.

The intrinsic reorientation time of the bacteria ($\tau_\text{R}$) is difficult to measure experimentally. When a bacterium collides with the two-phase boundary, it may change direction due to local torques, masking the underlying rotation caused by thermal motion and flagellar unbundling. Thus, we allow the non-dimensional parameter $L^2 \zeta/(\tau_\text{R}k_\text{B}T )$ to vary and be determined by a non-linear regression against the experimental data. Likewise, we determine the parameter $\delta/L$ in the same regression. The interface width $\delta$ arises when we model the two-phase mixture as a potential field, and it represents the boundary layer between the DEX-rich and PEG-rich phases. This interface width can (in principle) vary in space and time, so we allow it to vary as well.

We solve Eqs.~\ref{eq:1DAdvectionDiffusion} and \ref{eq:partition} at various $F_\text{max}/F_\text{prop}$ ratios, and fit the parameters by comparing the predicted partitioning against the experimental measurements. For clarity, we convert the $F_\text{max}/F_\text{prop}$ into DEX concentrations when plotting Fig.~\ref{fig:partitioning result}, obtained from our experimental data of $F_\text{max}$ in Fig.~\ref{fig:force&energy curve}C. The sensitivity of the solution to different parameter choices is demonstrated using a sensitivity analysis shown in Fig.~S9.

As shown in Fig.~\ref{fig:partitioning result}, our theory predicts that DEX-rich partitioning increases as the maximum restoring force increases.
Similar to the experiments with motile bacteria, our theory curve also increases monotonically with DEX concentration from $0.57 \pm 0.14$ at  1.5 wt/wt\% to $0.99 \pm 0.07$ at 8.0 wt/wt\%, corresponding to experimental ratios of $0.58 \pm 0.10$ and $1.00$ respectively.
As the energy barrier (and $F_\text{max}$) grows, it becomes difficult for this bacterium to penetrate the interface using their propulsion force alone. 
At a steady state, more bacteria are trapped in the low energy region, \textit{i.e.}, DEX-rich phase, unless they make a very large Brownian hop into the PEG-rich region.
In contrast, when the energy barrier is weak, the motile bacteria easily penetrate the interface and evenly distribute over the two phases.

From the non-linear regression we estimate the two fitting parameters as $\log_{10}[L^2\zeta/(\tau_\text{R}k_BT)] = 8.0 \pm 2.0$ and $\log_{10}[\delta/L] = -2.6 \pm 1.8$, with the error representing the 95$\%$ confidence interval.
The large error in our parameter estimation arises from the large uncertainties in the squared power of the heterogeneous domain size ($L$; large heterogeneities in domain size and shapes shown in Figs. S7C and S8), the effect of the confined chamber on the drag coefficient ($\zeta$), and the reorientation time ($\tau_\text{R}$) as discussed earlier. 
The interface thickness ($\delta)$ also has a large uncertainty from the experimental difficulty of measuring the DEX/PEG interface thickness modeled as a potential in our theory.
As such, there are multiple sets of parameters that give identical partitioning, making it difficult to distinguish the individual effects of these variables and leading to large uncertainties in their estimation. 
Physically, an increase in the rotational relaxation time $\tau_\text{R}$ (lower tumbling rate) has a similar effect on partitioning as an increase in the interfacial thickness $\delta$ (because the two fitting parameters are anticorrelated). 
In our model, a larger interface means that a bacterium is experiencing the chemical affinity force for a longer amount of time, so it must remain oriented perpendicular to the interface for longer in order to cross from the PEG-rich phase into the DEX-rich phase. 
Decoupling these parameters would require additional measurements that capture the spatial distribution and orientation of the bacteria.

Our model and experiments have differences, such as assuming the two phases form a lamellar pattern with a flat interface. 
As seen in Fig.~\ref{fig:microscopy}, the ATPS has a highly curved interface, which can affect the distribution of bacteria.
Moreover, the actual maximum force that freely motile bacteria overcome when crossing the interface is challenging to quantify precisely. 
Nevertheless, $F_\text{max}$ measured in the optical tweezers experiment is a reporter of the barrier force, as the contract angles of the dragged bacterium and the freely swimming bacterium were the same (See the inset of Fig.~S5A(5), inset of Fig.~S6(3) and Movie~S4). Note that the interfacial deformation and tension contribute to the force considerably, which is supported by the fact that $F_\text{max}$ increases as the DEX concentration and the resulting interfacial tension increase~\cite{Atefi2014}.
Finally, \textit{B. subtilis} cells vary greatly in their body lengths, propulsive force, and tumbling rates; the bacteria's propulsive force and the PEG-rich domain size heterogeneities are shown in Supplemental Fig. S7. 
This heterogeneity means that some bacteria are more likely to cross the interface than others, potentially enabling the ATPS to partition and sort the cells according to these properties.
These factors will all increase the uncertainty of both the theoretical model and parameter estimation.
Furthermore, as discussed earlier, while we do not expect the observed viscosity differences between the DEX-rich and PEG-rich regions to be large enough to explain the preferential partitioning present in our system, MIPS may contribute by slightly enhancing the partitioning effect as well.

In conclusion, we studied the competition between motility and two-phase boundaries on the biased partitioning of motile \textit{B. subtilis} using experiment and theory.
The motility of \textit{B. subtilis} helps to distribute the bacteria across both DEX-rich and PEG-rich phases by providing a large propulsive force, $F_\text{prop}\approx$ 10pN, to overcome the maximum interfacial force. 
We found that the chemical affinity of the \textit{B. subtilis} cell wall with the DEX-rich phase generates interfacial forces on the bacterium with magnitudes spanning from $F_\text{max} = $ 2pN to 10pN, depending on the DEX composition.

\vspace{20pt}

\begin{acknowledgments} 
J.C. and J.J. acknowledge the financial support from the National Research Foundation of Korea under NRF-2020R1A4A1019140 and NRF-2021R1A2C101116312.
R.J.M. is supported by the National Institute of Health (Grant No. 2023ER21800).
K.H.C., K.J.M., and S.C.T. are supported by the Air Force Office of Scientific Research under award number FA9550-21-1-0287.
K.J.M. is also supported by the National Science Foundation Graduate Research Fellowship under Grant No.~2139319.
S.C.T. is also supported by the Packard Fellowship in Science and Engineering.
\end{acknowledgments}

\bibliography{reference.bib}

\end{document}